\documentclass[prd,aps,floats,twocolumn,nofootinbib]{revtex4-1}
\usepackage{slashed}
\usepackage{mathtools}
\usepackage{amsfonts}
\usepackage{amssymb}
\usepackage{amsmath}
\usepackage{epsfig}

\begin{document}

\newcommand{\m}[1]{\mathcal{#1}}
\newcommand{\nn}{\nonumber}
\newcommand{\ph}{\phantom}
\newcommand{\eps}{\epsilon}
\newcommand{\be}{\begin{equation}}
\newcommand{\ee}{\end{equation}}
\newcommand{\bea}{\begin{eqnarray}}
\newcommand{\eea}{\end{eqnarray}}
\newtheorem{conj}{Conjecture}

\newcommand{\plk}{\mathfrak{h}}
\newcommand{\bb}{\bar b}


\title{Higgs Mechanism for the Ashtekar Self-Dual Connection}
\date{\today}


\author{Bruno Alexandre$^1$}
\email{bruno.alexandre20@imperial.ac.uk}

\affiliation{$^1$Theoretical Physics Group, The Blackett Laboratory, Imperial College, Prince Consort Rd., London, SW7 2BZ, United Kingdom}

\begin{abstract}

We introduce the Higgs mechanism for the self-dual spin connection (also known as the Ashtekar connection), using the Pleba\'{n}ski formulation of gravity. We develop our formalism within the framework of the chiral action and derive the equations of motion of the theory. One particular test model is explored: since anisotropy is an intrinsic property of the theory, a modified version of the spatially flat Bianchi I model with two different scale factors is considered. We apply our formalism and derive the Friedmann equations which regulate the scale factors and the Higgs field. We also present a Proca-like term for the connection, which when reduced to minisuperspace with a positive $\Lambda$ yields a De Sitter universe with an effective cosmological constant that depends on the mass of the gauge fields. We finally investigate the effect of these mass terms on gravitational waves and find that the wave equation remains unchanged relatively to GR; however the Weyl tensor is scaled by a constant which depends on the mass of the connection components.

\end{abstract}

\maketitle

\section{Introduction}

The search for an alternative theory to General Relativity (GR) has been going on for quite some time. One of the motivations for this quest is the fact that GR is not a complete theory and it has some drawbacks, among them the cosmological constant problem \cite{Lombriser,Padilla,Weinberg}, singularities and black holes and the lack of compatibility with a Quantum theory \cite{Isham,Kuchar}. 
In GR the gravitational force is predicted to be mediated by a massless spin-$2$ particle, the graviton. If we allow this field to be massive \cite{Crham} we introduce modifications to the gravitational interactions and we can explore pertinent questions such as the nature of dark energy and the accelerated expansion of the universe beyond the scope of GR. Two main points of concern rise when dealing with massive gravitons: 
they carry $5$ degrees of freedom as opposed to the $2$ carried out by the massless ones and the fact that most extensions of the Fierz-Pauli action will have present a Boulware–Deser (BD) ghost. The first one can be resolved through the Vainshtein mechanism, the later can be avoided recurring to other formulations of massive gravity such as soft massive gravity, new massive gravity and dRGT ghost free massive gravity.

Another possibility to investigate modified gravity is to consider other formulations of gravity as a starting point. Rather than beginning with the Einstein-Hilbert (HB) formalism, which encodes gravity in the metric field, one can build the action in terms of self dual $2$-forms which can be related to the metric by setting additional constraint equations. This is the Pleba\'{n}ski formulation of gravity \cite{Plebanski}, where one has a first order Lagrangian for GR that contains only the self-dual (SD) part of the spin connection. This theory is vastly explained in \cite{PracticalIntro,KirillBook} and some other works have been carried out exploring it \cite{capo,krasnov1}. In particular, an interesting extension of Pleba\'{n}ski formalism is the ``pure connection'' formulation \cite{krasnov2,ElliotSteffen,ElliotSteffen2} where the two-form fields representing the metric are integrated out of the action, leaving only the connection to the metric as the fundamental field in the theory.

Within this formalism, a natural way to consider modified gravity is to promote the cosmological constant to a function of the Lagrange multiplier present in the theory \cite{simone}. Therefore one could ask what happens if $\Lambda$ is a function of some other variable present in the theory, for instance a scalar field. In this paper we will consider a rather specific version of this proposal: $\Lambda$ is replaced in the gravitional action by the Higgs scalar field potential.
The Higgs mechanism \cite{higgs} involves the introduction of a scalar field, commonly referred to as the Higgs boson, that interacts with particles within the Standard Model (SM). This interaction results in the acquisition of mass by these particles. Hence one can wonder if such mechanism can somehow be used to ``give'' mass to the gravity gauge fields. In order to address this question, we need to consider an alternative way of describing GR. Thus the Pleba\'{n}ski formulation of gravity mentioned above seems to be a good option as the SD connection is the ideal candidate to become massive in this theory through the Higgs mechanism. This differs from the work done in \cite{moffat} as in that case the coupling of the Higgs is done through the $SO(3,1)$ spin connection from the Einstein-Cartan formalism, which changes the nature of the problem.

The aim of this paper is to implement and explore the Higgs mechanism in Pleba\'{n}ski gravity, which is done by coupling an $SU(2)$ complex scalar field to the Ashtekar SD connection in the gravitational action. We briefly introduce the necessary mathematical tools for the theory and further establish the whole formalism, from the action to the equations of motion. We additionally consider a case where the connection acquires mass from a Proca-like term rather than through the spontaneous symmetry breaking of the Higgs field.
The paper is structured in the following way. In section \ref{secplebanski} we introduce the chiral Pleba\'{n}ski formulation of GR. We write down the action, present all the important physical quantities and derive the equations of motion. In section \ref{sechig} we present the Higgs mechanism in this formalism and show how the broken symmetry originates the mass terms for the spin connection. We then apply this to a modified Bianchi I model. We then investigate the alternative version of having a Proca-like term for the connection in section \ref{secproca} and in section \ref{secgw} we analyse the consequence of this mass terms on gravitational waves. We conclude with some final remarks on current and future work in section \ref{secconc}.

\section{Pleba\'{n}ski formulation}
\label{secplebanski}

The chiral Pleba\'{n}ski formulation of gravity encodes the metric into complex self-dual 2-forms and gravity is mediated by a complex $SU(2)$ connection $A^i$ \cite{Plebanski,PracticalIntro,KirillBook}. An action for GR with cosmological constant $\Lambda$ is given by
\begin{equation}S=\frac{i}{\ell_P^2}\int \Sigma^i F_i -\frac{1}{2}M_{ij}\Sigma^i\Sigma^j+\frac{1}{2}\omega({\rm tr}\,M-\Lambda),
\label{eqplebac}
\end{equation}
where
\begin{eqnarray}
F^i=dA^i+\frac{1}{2}\epsilon^{i}_{\,\,jk}A^{j}A^k
\label{eqF}
\end{eqnarray}
is the curvature complex 2-forms of the connection $A^i$, $\Sigma^i$ are 2-forms, $M_{ij}$ is a symmetric $3\times 3$ matrix, $\omega$ is a complex $4$-form and $\ell_P^2=8\pi G$ is the Planck length.

The equations of motion are obtained varying the action with respect to $A^i$, $\Sigma^i$ and $M^{ij}$, respectively: 
\begin{align}
D_A\Sigma^i &=d\Sigma^i+\epsilon^{i}_{\,\,jk}A^{j}\Sigma^k=0\, \label{PlebEq1} \\ F^i &=M^{ij}\Sigma_j\, \\ \Sigma^i\Sigma^j & = \delta^{ij}\omega, \label{PlebEq2}
\end{align}
together with the constraint ${\rm tr}\, M=\Lambda$ coming from variation with respect to $\omega$. Solutions to these equations represent solutions of complex general relativity. Lorentzian solutions are obtained by imposing reality conditions
\begin{equation}
{\rm Re}(\Sigma^i\Sigma_i)=\Sigma^i\bar{\Sigma}^j=0
\end{equation}
whereas Euclidean solutions correspond to real $\Sigma^i$. 
For non-zero $\omega$ one of the possible solutions for $\Sigma^i$ \cite{ElliotSteffen} is given by
\begin{eqnarray}
\Sigma^i=ie^0e^i-\frac{1}{2}\epsilon^{i}_{\,\,jk}e^je^k,
\label{eqsd}
\end{eqnarray}
where $e^i$ is the tetrad of a given metric. $\bar{\Sigma}^j=-(\Sigma^{i})^*$ is the anti-self-dual (ASD) $2$-form. A metric can be recovered from the $\Sigma^i$ using the Urbantke formula \cite{Urbantke}
\begin{align}
g_{\mu\nu}\,\epsilon_{\Sigma}&=-\frac{i}{6}\epsilon_{ijk}\,i_\mu\Sigma^i\,i_\nu\Sigma^j\,\Sigma^k\,,
\\\epsilon_{\Sigma}&=\frac{i}{6}\Sigma^i\Sigma_i=\sqrt{-g}d^4x
\end{align}
where $i_\mu\Sigma^i$ is a 1-form defined by $(i_\mu\Sigma^i)_\nu=\Sigma^i_{\mu\nu}$.

\section{Gravity with a Higgs Field}
\label{sechig}
We now couple the complex doublet of $SU(2)$ Higgs field $\phi^i$ to the above gravity action and replace the cosmological constant by the Higgs potential:
\begin{eqnarray}
S=&\frac{i}{\ell_P^2}\int \Sigma^i F_i -\frac{1}{2}M_{ij}\Sigma^i\Sigma^j+\frac{1}{2}\omega[{\rm tr}\,M-\ell_P^2V(\phi)]+ \nonumber \\
& +\int\frac{1}{2} D_A\phi^i\ast D_A\phi_i,
\label{eqah}
\end{eqnarray}
with 
\begin{eqnarray}
    V(\phi)=\frac{\lambda}{4}(\phi_i^2-v^2)^2,
\end{eqnarray}
where $\lambda$ is a coupling constant and $v$ is the vacuum expectation value (vev) of the Higgs field. The Hodge dual can be defined for a general $p$-form $A$ in $n$ dimensions using the Urbantke metric as:
\begin{eqnarray}
    \star A=\frac{1}{(n-p)!p!}\epsilon_{\mu_1...\mu_{n-p}\nu_1...\nu_p}A^{\nu_1...\nu_p}dx^{\mu_1}...dx^{\mu_{n-p}}
\end{eqnarray}
and $\epsilon_{\mu_1...\mu_n}=\sqrt{-g}\varepsilon_{\mu_1...\mu_n}$ is the Levi-Civita tensor. $\varepsilon_{\mu_1...\mu_n}$ is the Levi-Civita symbol and we use the conventions $\varepsilon_{0123}=1$ and $\varepsilon^{0123}=-1$, which means we have $\epsilon^{\mu_1...\mu_n}=\frac{\varepsilon^{\mu_1...\mu_n}}{\sqrt{-g}}$.

The equations of motion for this new action are:
\begin{eqnarray}
   & D_A\Sigma^i=i\ell_P^2\epsilon^{i}_{\,\,jk}\phi^j\ast D_A\phi^k \\
& F^i=M^{ij}\Sigma^j-\ell_P^2\tau^i \label{eqmF} \\
& {\rm tr}\, M=\ell_P^2 V(\phi) \\
& D_A\ast D_A\phi_i=-\frac{iV'}{6\ell_P^2}\Sigma^j\Sigma_j,
\end{eqnarray}
where
\begin{eqnarray}
    \tau_i=\frac{\delta S_{D_A\phi}}{i\delta\Sigma^i}
    \label{eqtau}
\end{eqnarray}
is the 2-form stress energy tensor associated with the Hodge dual term of the scalar field.

We now expand around the vacuum state as \cite{unigauge}
\begin{eqnarray}
    \phi^i=\phi_0^i+\varphi^i
\end{eqnarray}
with
$\phi^i_0=v\delta^i_3$.
Considering the covariant derivative of the scalar field
\begin{eqnarray}
D_A\phi^i=d\phi^i+\epsilon^{i}_{\,\,jk}A^j\phi^k,
\end{eqnarray}
the kinetic energy term takes the form
\begin{eqnarray}
      D_A\phi^i\ast D_A\phi_i=d\varphi_3\ast d\varphi_3+v^2(A^1\ast A^1+A^2\ast A^2),
\end{eqnarray}
where we kept only quadratic terms in the fields and we have eliminated the terms of the form $v\epsilon_{ij3}d\varphi^i\ast A^j$ with a gauge transformation, for instance the unitary gauge from electroweak theory \cite{unigauge}. Within this approximation the potential $V(\phi)$ becomes $V(\varphi_3)=v^2\lambda\varphi_3^2$. 
Replacing this back in the action (\ref{eqah}) we obtain:
\begin{eqnarray}
S=&\frac{i}{\ell_P^2}\int \Sigma^i F_i -\frac{1}{2}M_{ij}\Sigma^i\Sigma^j+\frac{1}{2}\omega[{\rm tr}\,M-\ell_P^2V(\varphi_3)]+ \nonumber \\
& +\int\frac{1}{2} d\varphi_3\ast d\varphi_3+\frac{1}{2}\int v^2(A^1\ast A^1+A^2\ast A^2).
\label{eqhaction}
\end{eqnarray}
One can now explicitly see the effect of the Higgs mechanism: two of the gravity gauge fields have a mass term and the mass is determined by the vev! 
The only two equations of motion that are affected by the expansion around the vev become:
\begin{eqnarray}
    & D_A\Sigma^i=i\ell_P^2v^2(\ast A^1\delta^{i1}+\ast A^2\delta^{i2}) 
    \label{eqds1}
\end{eqnarray}
and
\begin{eqnarray}
   & d\ast d\varphi_3=-\frac{i}{6}\Sigma^i\Sigma_iV' \\
   & \Rightarrow \Box \varphi_3=V'.
\label{eqfi}
\end{eqnarray}
Hence, we see clearly now from equation (\ref{eqds1}) that there is a torsion contribution term from the massive gauge fields and equation (\ref{eqfi}) is simply the Klein-Gordon equation for a scalar field with a potential. 

\subsection{Modified Bianchi I Model}
Upon examining action (\ref{eqhaction}), it becomes clear that the mass terms of the two gauge fields introduce anisotropy into the theory. As a result, solutions cannot be sought within homogeneous and isotropic models, such as the FLRW spacetime, due to an inherent inconsistency between the connection and the area form. Therefore we must look for anisotropic solutions like the Bianchi I model, or at least a simplified version of it. 

In this section we consider a modified Bianchi I model where there are only two scale factors and consequently only one preferred direction.
The metric for this universe is
\begin{eqnarray}
    ds^2=-dt^2+a^2(t)(dx^2+dy^2)+b^2(t)dz^2,
\end{eqnarray}
where $t, x, y, z$ are the comoving coordinates $x^0, x^1, x^2, x^3$, $a(t)$ is the scale factor in the $x,y$ direction and $b(t)$ corresponds to the $z$ direction. From the metric one can obtain the self-dual area form given by equation (\ref{eqsd}), which yields:
\begin{eqnarray}
   && \Sigma^1=iadtdx^1-abdx^2dx^3 \\
   && \Sigma^2=iadtdx^2-abdx^3dx^1 \\
   && \Sigma^3=ibdtdx^3-a^2dx^1dx^2.
\end{eqnarray}
We then plug this in equation (\ref{eqds1}) in order to solve for the SD connection $A^i$:
\begin{eqnarray}
   && A^1=i\dot adx^1 \\
   && A^2=i\dot adx^2 \\
   && A^3=i(\dot b-\ell_P^2v^2b\dot a/a)dx^3. 
\end{eqnarray}
It is interesting to notice  that comparing to GR the two massive gauge fields stay unchanged and only the massless field gets a contribution related to the vev, which involves both scale factors. Using now (\ref{eqF}) we get 
\begin{eqnarray}
F^i=&&\left[\frac{\Ddot{a}}{a}+\frac{\dot a\dot b}{ab}-\ell_P^2v^2\frac{\dot a^2}{a^2}\right]\frac{\Sigma^i}{2}+\\
    && +\left[\frac{\Ddot{a}}{a}-\frac{\dot a\dot b}{ab}+\ell_P^2v^2\frac{\dot a^2}{a^2}\right]\frac{\bar\Sigma^i}{2} \\
    F^3=&& \left[\frac{\Ddot{b}}{b}+\frac{\dot a^2}{a^2}-\ell_P^2v^2\left(\frac{\Ddot a}{a}+\frac{\dot a\dot b}{ab}-\frac{\dot a^2}{a^2}\right)\right]\frac{\Sigma^3}{2}+\\
    && +\left[\frac{\Ddot{b}}{b}-\frac{\dot a^2}{a^2}-\ell_P^2v^2\left(\frac{\Ddot a}{a}+\frac{\dot a\dot b}{ab}-\frac{\dot a^2}{a^2}\right)\right]\frac{\bar\Sigma^3}{2} ,
\end{eqnarray}
 with $i=1,2$. 
 
 The only quantity left to calculate from equation (\ref{eqmF}) is the $2$-form stress-energy tensor. It is not clear from the existent literature how to compute equation (\ref{eqtau}), therefore we use the following formula from \cite{KirillBook} to ``translate'' the stress tensor from the EH to the Pleba\'{n}ski formalism:
\begin{eqnarray}
\tau^i=\frac{T}{12}\Sigma^i-\frac{1}{2}\Sigma^{i\,\,\,\,\rho}_{\,\,\mu}\Tilde{T}_{\rho\nu}dx^\mu dx^\nu,
\end{eqnarray}
where $\Tilde{T}_{\mu\nu}=T_{\mu\nu}-\frac{1}{4}g_{\mu\nu}T$, $T=g^{\mu\nu}T_{\mu\nu}$ and $T_{\mu\nu}$ is the stress-energy tensor defined as:
\begin{eqnarray}
    T_{\mu\nu}=-\frac{2}{\sqrt{-g}}\frac{\delta S_{\text{matter}}}{\delta g^{\mu\nu}}.
    \label{eqst}
\end{eqnarray}
For the case of action (\ref{eqhaction}) the terms that contribute to $T_{\mu\nu}$ are the ones with the Hodge dual operator (which correspond to inner products and therefore carry a metric) and so one arrives at:
\begin{widetext}
    \begin{eqnarray}
T_{\mu\nu}=v^2(A_\mu^1A_\nu^1+A_\mu^2A_\nu^2)-g_{\mu\nu}\frac{v^2}{2}(A_\alpha^1A^{1\alpha}+A_\alpha^2A^{2\alpha})+\partial_\mu\varphi_3\partial_\nu\varphi_3-\frac{1}{2}g_{\mu\nu}(\partial\varphi_3)^2.
    \end{eqnarray}
\end{widetext}
The trace and non-zero traceless parts of this tensor are respectively:
\begin{eqnarray}
    T=2v^2\frac{\dot a^2}{a^2}+\dot\varphi_3^2
    \label{eqsttr}
\end{eqnarray}
and 
\begin{eqnarray}
   && \Tilde{T}_{00}=\frac{3}{4}\dot\varphi_3^2-\frac{v^2\dot a^2}{2a^2} \\
    && \Tilde{T}_{11}=\Tilde{T}_{22}=\frac{a^2}{4}\dot\varphi_3^2-\frac{v^2\dot a^2}{2} \\
     && \Tilde{T}_{33}=\frac{b^2}{4}\dot\varphi_3^2+\frac{v^2\dot a^2b^2}{2a^2}. 
\end{eqnarray}
This is everything we need to compute $\tau^i$ which can be explicitly written as

\begin{eqnarray}
\tau^1=&&\frac{T}{12}\Sigma^1-\frac{1}{2}\left(-ia\Tilde{T}_{00}+\frac{i}{a}\Tilde{T}_{11}\right)dtdx^1+ \nonumber \\
&&+\frac{1}{2}\left(\frac{b}{a}\Tilde{T}_{22}+\frac{a}{b}\Tilde{T}_{33}\right)dx^2dx^3 \\
\tau^2=&&\frac{T}{12}\Sigma^2-\frac{1}{2}\left(-ia\Tilde{T}_{00}+\frac{i}{a}\Tilde{T}_{22}\right)dtdx^2+ \nonumber\\
&&+\frac{1}{2}\left(\frac{b}{a}\Tilde{T}_{11}+\frac{a}{b}\Tilde{T}_{33}\right)dx^3dx^1 \\
\tau^3=&&\frac{T}{12}\Sigma^3-\frac{1}{2}\left(-ib\Tilde{T}_{00}+\frac{i}{b}\Tilde{T}_{33}\right)dtdx^3+ \nonumber\\
&&+\frac{1}{2}\left(\Tilde{T}_{11}+\Tilde{T}_{22}\right)dx^1dx^2.
\end{eqnarray}
Thus we finally get
\begin{eqnarray}
    \tau^i=\frac{1}{12}\left(2v^2\frac{\dot a^2}{a^2}+\dot\varphi_3^2\right)\Sigma^i+\frac{\dot\varphi_3^2}{4}\bar\Sigma^i, \label{eqtaui}
\end{eqnarray}
for $i=1,2$ and
\begin{eqnarray}
    \tau^3=\frac{1}{12}\left(2v^2\frac{\dot a^2}{a^2}+\dot\varphi_3^2\right)\Sigma^3-\left(\frac{v^2\dot a^2}{2a^2}-\frac{\dot\varphi_3^2}{4}\right)\bar\Sigma^3, \label{eqtau3}
\end{eqnarray}
for $i=3$.
We can then replace all the components of $F^i$ and $\tau^i$ in (\ref{eqmF}) to get four equations, two corresponding to the components $i=1$ and $i=2$
\begin{eqnarray}
    & \frac{\Ddot{a}}{2a}+\frac{\dot a\dot b}{2ab}-\ell_P^2v^2\frac{\dot a^2}{2a^2}=M_{ii}-\frac{\ell_P^2}{12}\left(2v^2\frac{\dot a^2}{a^2}+\dot\varphi_3^2\right) \label{eqsd1} \\
& \frac{\Ddot{a}}{2a}-\frac{\dot a\dot b}{2ab}+\ell_P^2v^2\frac{\dot a^2}{2a^2}=-\frac{\ell_P^2}{4}\dot\varphi_3^2, \label{eqasd1}
\end{eqnarray}
where there is no sum over $i$, and two equations corresponding to $i=3$: 
\begin{eqnarray}
    & \frac{\Ddot{b}}{2b}+\frac{\dot a^2}{2a^2}-\frac{\ell_P^2v^2}{2}\left(\frac{\Ddot a}{a}+\frac{\dot a\dot b}{ab}-\frac{\dot a^2}{a^2}\right)=M_{33}- \nonumber \\
    & -\frac{\ell_P^2}{12}\left(\frac{2v^2\dot a^2}{a^2}+\dot\varphi_3^2\right) \label{eqsd2} \\
& \frac{\Ddot{b}}{2b}-\frac{\dot a^2}{2a^2}-\frac{\ell_P^2v^2}{2}\left(\frac{\Ddot a}{a}+\frac{\dot a\dot b}{ab}-\frac{\dot a^2}{a^2}\right)= \nonumber \\ 
& =\ell_P^2\left(\frac{v^2\dot a^2}{2a^2}-\frac{\dot\varphi_3^2}{4}\right). \label{eqasd2}
\end{eqnarray}

The first Friedmann equation for this universe can be obtained by combining (\ref{eqsd1})$-$(\ref{eqasd1}) and (\ref{eqsd2})$-$(\ref{eqasd2}) together with the constraint ${\rm tr}\, M=\ell_P^2V$:
\begin{eqnarray}
    2\frac{\dot a\dot b}{ab}+\frac{\dot a^2}{a^2}(1-\ell_P^2v^2)=\ell_P^2\left(\frac{\dot\varphi_3^2}{2}+V\right).
    \label{eqF1}
\end{eqnarray}
For the second Friedmann equation we do (\ref{eqsd1})$+$(\ref{eqasd1}) and (\ref{eqsd2})$+$(\ref{eqasd2}) together with the constraint ${\rm tr}\, M=\ell_P^2V$:
\begin{eqnarray}
    2\frac{\Ddot a}{a}+\frac{\Ddot b}{b}-\ell_P^2v^2\left(\frac{\Ddot a}{a}+\frac{\dot a\dot b}{ab}-\frac{\dot a^2}{a^2}\right)=\ell_P^2\left(-\dot\varphi_3^2+V\right).
    \label{eqF2}
\end{eqnarray}
One can see that setting $v=0$ recovers the usual GR Friedmann equations.
The remaining equation to consider is the equation of motion for the scalar field (\ref{eqfi}) in this spacetime, which takes the form:
\begin{eqnarray}
    \Ddot{\varphi}_3+\left(\frac{2\dot a}{a}+\frac{\dot b}{b}\right)\dot\varphi_3+V'(\varphi_3)=0.
    \label{eqmofi}
\end{eqnarray}
Solving together (\ref{eqF1}), (\ref{eqF2}) and (\ref{eqmofi}) fully determines $a(t)$, $b(t)$ and $\varphi_3(t)$ and consequently, all the dynamics of spacetime for this theory.

\section{Proca Term for the Ashtekar Connection}
\label{secproca}

So far we have considered that the Ashtekar connection acquires mass through the Higgs mechanism, resulting in only two of its three components becoming massive. However, what if we want all components to acquire mass? While this cannot be achieved using the same mechanism, we can always add a Proca term to the action  (\ref{eqplebac}), corresponding to having three gauge fields with mass $m$. Hence this reads:
\begin{eqnarray}
S=&\frac{i}{\ell_P^2}\int \Sigma^i F_i -\frac{1}{2}M_{ij}\Sigma^i\Sigma^j+\frac{1}{2}\omega[{\rm tr}\,M-\Lambda]+ \nonumber \\
& +\frac{1}{2}\int m^2A^i\ast A^i.
\label{eqproaction}
\end{eqnarray}
The first change we get is in the torsion equation of motion, for which we now obtain 
\begin{eqnarray}
    & D_A\Sigma^i=i\ell_P^2m^2\ast A^i.
    \label{eqds}
\end{eqnarray}
We can see that contrary to what happens in the Higgs mechanism case, all the components of the self-dual connection obey the same equation. This indicates that we are now dealing with an isotropic theory and hence we can seek a general solution for an FLRW universe:
\begin{eqnarray}
    \Sigma^i=iadtdx^i-\frac{1}{2}a^2\epsilon^{i}_{\,\,jk}dx^jdx^k.
\end{eqnarray}
Plugging this in (\ref{eqds}) yields
\begin{eqnarray}
    A^i=\frac{2i\dot a}{2+\ell_P^2m^2}dx^i
\end{eqnarray}
and the associated curvature is 
\begin{eqnarray}
F^i=&&\left(\frac{\Ddot{a}}{a(2+\ell_P^2m^2)}+\frac{2\dot a^2}{a^2(2+\ell_P^2m^2)^2}\right)\Sigma^i+ \nonumber \\
&&+\left(\frac{\Ddot{a}}{a(2+\ell_P^2m^2)}-\frac{2\dot a^2}{a^2(2+\ell_P^2m^2)^2}\right)\bar{\Sigma}^i.
\end{eqnarray}
The structure of the remaining equations of motion stays unchanged, nonetheless the stress energy tensor takes the form
\begin{eqnarray}
    T_{\mu\nu}=m^2A^i_\mu A^i_\nu-g_{\mu\nu}\frac{m^2}{2}A^i_\alpha A^{i\alpha}
\end{eqnarray}
and hence the $2$-form stress tensor is now: 
\begin{eqnarray}
    \tau^i=\frac{m^2}{(2+\ell_P^2m^2)^2}\frac{\dot a^2}{a^2}(\Sigma^i-\bar\Sigma^i).
\end{eqnarray}
Based on the work done in the previous sections it is straightforward to derive the final equations of motion using the self-dual and anti-self-dual parts of (\ref{eqmF}), which together with ${\rm tr}\, M=\Lambda$ yield
\begin{eqnarray}
    \Ddot a=\frac{\dot a^2}{a}
\end{eqnarray}
and
\begin{eqnarray}
    \frac{\dot a^2}{a^2}=\frac{\Lambda_{\text{eff}}}{3}.
\end{eqnarray}
This is nothing but the usual equation for the scale factor in FLRW spacetime in GR but with an effective cosmological constant given by
\begin{eqnarray}
\Lambda_{\text{eff}}=\frac{2+m^2\ell_P^2}{2}\Lambda.
\label{eqlef}
\end{eqnarray}
The solution then depends on the mass of the gauge fields and on $\Lambda$. We have $\Lambda_{\text{eff}}>0$ and therefore a real scale factor, representing a dS spacetime.







\section{Gravitational Waves in Pleba\'{n}ski Gravity}
\label{secgw}

In this section we introduce gravitational waves in the Pleba\'{n}ski formalism. This is necessary if one wants to study the effect of having mass terms for the $SU(2)$ connection on the gravitational wave equations. We are going to focus on the case introduced in the previous section where all the components of the connection are massive and let us also set $\Lambda=0$. 

Therefore, the background is described by the Minkowski metric (here represented by the area form $\Sigma_0^i$) and the tensorial perturbations corresponding to gravitational waves are 
\begin{eqnarray}
\Sigma^i=\Sigma_0^i+\delta\Sigma^i=\Sigma_0^i+h^{ij}\bar{\Sigma}_0^j,
\end{eqnarray}
where $h^{ij}$ is symmetric, traceless and obeys $h^{ij}_{\,\,\,,i}=0$ and 
\begin{eqnarray}
    \Sigma_0^i=idtdx^i-\frac{1}{2}\epsilon^{i}_{\,\,jk}dx^jdx^k.
\end{eqnarray}
The connection can be decomposed in the same way
\begin{eqnarray}
    A^i=A_0^i+\delta A^i=\delta A^i,
\end{eqnarray}
with $A_0^i=0$ as this is the connection associated with the vacuum.
The aim is to determine $\delta A^i$ in terms of $h^{ij}$, for which we use 
\begin{eqnarray}
&& d\Sigma^i+\epsilon^{i}_{\,\,jk}A^j\Sigma^k=i\ell_P^2m^2\ast A^i, \\
 \Leftrightarrow && d\delta\Sigma^i+\epsilon^{i}_{\,\,jk}\delta A^j\Sigma_0^k=i\ell_P^2m^2\ast \delta A^i.
\end{eqnarray}
Solving this equation of motion as in the previous sections results in:
\begin{eqnarray}
    \delta A^i=\frac{1}{1+\ell_P^2m^2}(i\dot h^{ij}-h^{ik}_{\,\,\,,l}\epsilon^{j}_{\,\,kl})dx^j,
\end{eqnarray}
where the dot represents a derivative with respect to time. The curvature is obtained through 
\begin{eqnarray}
&& F^i=dA^i+\frac{1}{2}\epsilon^{i}_{\,\,jk}A^{j}A^k \\
\Leftrightarrow
&& F^i=F_0^i+\delta F^i=d\delta A^i
\end{eqnarray}
where $F_0^i=0$ and $\delta F^i$ can be decomposed as:
\begin{eqnarray}
\delta F^i=\psi^{ij}\Sigma^j+\bar\psi^{ij}\bar{\Sigma}^{j}.
\end{eqnarray}
The anti-self dual part must be $0$ and therefore we get the wave equation:
\begin{eqnarray}
\Box h^{ij}=0,
\label{eqwe}
\end{eqnarray}
whose solutions are plane waves.
The self-dual component leads to the Weyl tensor part of the curvature yielding:
\begin{eqnarray}
\psi^{ij}=\frac{1}{1+\ell_P^2m^2}\left(\frac{1}{2}\Ddot{h}^{ij}+i\epsilon^{jkl}\dot h^{ik}_{\,\,\,,l}+\frac{1}{2}\nabla^2h^{ij}\right),
\end{eqnarray}
which is traceless as it should since ${\rm tr}\,M=0$.

We can then see that the nature of the gravitational wave equations are not altered by the presence of the mass terms in the theory as equation (\ref{eqwe}) is the usual plane wave equation for $h^{ij}$. Nevertheless the Weyl tensor gets an additional pre-factor that depends on the mass $m$ of the connection.

\section{Conclusions and Outlook}
\label{secconc}

In this paper we introduced the Higgs mechanism for the Pleba\'{n}ski formulation of gravity. This was done by adding an $SU(2)$ scalar field to the theory that couples with the self-dual spin connection and by replacing the cosmological constant with the Higgs field potential. This resulted in having two out of the three components of the gauge field becoming massive, which altered the equations of motion relatively to GR, bringing a mass dependent torsion into play and introducing a natural anisotropy in the theory. In order to explore further this proposal we were forced to leave out the usual MSS solution and considered a non-isotropic model of the universe with two different scale factors, i.e. a simplified version of the Bianchi I model, for which we derived all the equations of motion. However we left them to be solved in future work as numerical methods are required to do so. Nevertheless it is interesting to notice that, compared to GR, the massless component of the connection is the one that gets an extra term proportional to the vev squared.

We also considered a Proca-like term for the Ashtekar connection and therefore all its components became massive. When reduced to MSS with a cosmological constant $\Lambda$ this led to the usual dS universe solution with an effective cosmological constant which depends on the mass of the connection. Moreover, we applied this model to gravitational waves with tensorial perturbations in the Pleba\'{n}ski formalism and we found that for a Minkowski background the gravitational wave equation stays unchanged and only the Weyl tensor gets scaled by a constant also dependent on the mass of the connection.

Further work would involve numerically solving the equations to obtain the scale factors for the modified Bianchi I model, as mentioned above, and even generalizing for a Bianchi I model which contains three different scale factors. It would also be interesting to study gravitational waves in the formulation with the Higgs field as this differs from the Proca-like term since one of the gauge fields is massless (which implies an anisotropic background for the gravitational waves) and there is a scalar field involved. Finally, one could try to build the appropriate formalism to derive the $2$-form stress-energy tensor directly from the Pleba\'{n}ski gravitational action by varying the Hodge dual terms with respect to the self dual $2$-form.

\acknowledgments
We thank Jo\~{a}o Magueijo, Steffen Gielen and Raymond Isichei for helpfull discussions and comments.
This work was supported by the FCT Grant https://doi.org/10.54499/2021.05694.BD. 


\end{document}